# SCALING LAWS OF FREE MAGNETIC ENERGY STORED IN A SOLAR EMERGING FLUX REGION

T. Magara

*Dept. of Astronomy and Space Science, Kyung Hee University,*
*1732 Deogyeong-daero, Giheung-gu, Yongin, Gyeonggi-do, 446-701, Republic of Korea*
*magara@khu.ac.kr*

**Abstract**

This Letter reports scaling laws of free magnetic energy stored in a solar emerging flux region which is a key to understanding the energetics of solar active phenomena such as solar flares and coronal mass ejections. By performing 3-dimensional magnetohydrodynamic simulations that reproduce several emerging flux regions of different magnetic configurations, we derive power law relationships among emerged magnetic flux, free magnetic energy and relative magnetic helicity in these emerging flux regions. Since magnetic flux is an observable quantity, the scaling law between magnetic flux and free magnetic energy may give a way to estimate invisible free magnetic energy responsible for solar active phenomena.

## 1. Introduction

While the Sun usually appears to be quiet in white light, the Sun is actually full of active phenomena such as solar flares (Shibata & Magara 2011) and coronal mass ejections (Chen 2011); they are now observed on a daily basis. These active phenomena often occur in an emerging flux region with intense magnetic flux in it, which is called an *active region*. Since they are observed as one of the largest energy-release events on the Sun, the energetics of those active phenomena has been an important target of research in the solar physics. Some of them even affect the environment of the earth; in an extreme case they damage telecommunication and power supplies. This makes it important to investigate the energetics of solar active phenomena and their impact on the earth, which has emerged as the space weather.

When we study the energetics of solar active phenomena, it is important to estimate free magnetic energy required to produce an energy-release event. The free energy is defined as an excess over the potential energy that is the lowest energy taken by a magnetic structure formed on the Sun. Since it is still difficult to derive the overall configuration of a magnetic structure only by observations, especially in the solar corona where most of huge active phenomena originate, we cannot directly calculate free magnetic energy; instead we have to use observable quantities to estimate free magnetic energy.

One of the ways to estimate free magnetic energy is to reconstruct a coronal magnetic structure from an observed photospheric magnetic field. This has become useful when not only the vertical component of a photospheric field but also its transverse components are available so that by combining these components we can reconstruct a magnetic structure containing free magnetic energy in the corona. Reconstruction of a coronal magnetic structure based on the force-free approximation (e.g. Sturrock 1994) has well been investigated and developed (Wiegelmann & Sakurai 2012), and it has widely been applied to observed active regions to derive their magnetic configurations. There have been proposed various nonlinear force-free field (NLFFF) reconstruction methods and reviews of them with a quantitative comparison are given in Schrijver et al. (2006) and Metcalf et al. (2008). Some of recent work where NLFFF reconstruction methods are used for investigating observed active regions is found in Metcalf, Leka & Michey (2005), Régnier & Priest (2007), Inoue et al. (2012) and Sun et al. (2012), the last of which use an optimization code developed by Wiegelmann (Wiegelmann 2004). The validity of NLFFF reconstruction and how to improve it have also been studied (Metcalf et al. 1995; Wheatland & Metcalf 2006).

In this Letter we report scaling laws of free magnetic energy stored in an emerging flux region, which may give a supplementary way to estimate invisible free magnetic energy using observed magnetic flux. By performing a series of 3-dimensional magnetohydrodynamic (MHD) simulations that reproduce several emerging flux region configurations by bringing a magnetic flux tube of different twist to the solar atmosphere, we derive power law relationships among free magnetic energy, magnetic flux and relative magnetic helicity, the last of which is also an important quantity related to the activity of an emerging flux region (Jeong & Chae 2007; Lim et al. 2007). We also use a linear force-free field (LFFF) model to investigate power law relationships among these quantities. By comparing the simulations and the LFFF model we discuss the possible application of the scaling laws to an emerging flux region where a huge active phenomenon tends to occur.

## 2. Simulation setup and results

To perform a 3-dimensional simulation we solved ideal MHD equations, the details of which are described in Magara (2012). We use the Cartesian coordinates $(x,y,z)$ where the $x$- and $y$-axes form a horizontal plane corresponding to the solar surface ($z=0$), while the $z$-axis is directed upward. The present study uses a wider simulation domain $(-200,-200,-10)<(x,y,z)<(200,200,190)$ than that used in Magara (2012) so that long-term evolution can be investigated. The domain is discretized into those grids whose size is $(\Delta x, \Delta y, \Delta z)=(0.1,0.2,0.1)$ for $(-8,-12,-10)<(x,y,z)<(8,12,15)$ whereas it gradually increases up to $(4,4,4)$ as $|x|,|y|$ and $z$ increase. The total number of grids is $N_x \times N_y \times N_z = 371 \times 303 \times 353$. Inside this domain we put a background atmosphere stratified under uniform gravity. A magnetic flux tube with a finite radius $r_f = 2$, characterized by a Gold-Hoyle profile (Gold & Hoyle 1960) initially submerges under the solar surface. The axis of this flux tube is parallel to the $y$-axis, crossing the $z$-axis at $z = -4$. The flux tube and background atmosphere are initially in mechanical equilibrium, and to initiate a simulation we applied a velocity perturbation to the middle portion of the flux tube (we followed Equation (10) in Magara (2012), except for using $L = 200$ in the present study), which has then started to emerge in the shape of an $\Omega$ loop. The unit of normalization is given by $2\Lambda_{ph}$ (length), $c_{ph}$ (velocity), $\rho_{ph}$ (gas density), $\rho_{ph}c_{ph}^2$ (gas pressure), $T_{ph}$ (temperature) and $(\rho_{ph}c_{ph}^2)^{1/2}$ (magnetic field) where $\Lambda_{ph}$, $c_{ph}$, $\rho_{ph}$ and $T_{ph}$ represent the pressure scale height, adiabatic sound speed, gas density and temperature in the photosphere, respectively.

We ran a series of simulations by changing the twist parameter of field lines composing a Gold-Hoyle flux tube. In Magara (2012) we studies four cases where the twist parameter is given by 0.2, 0.35, 0.5 and 1; in the present study we add another case (twist parameter is 0.8) to them. Figures 1a-1c present snapshots of an emerging flux region in this newly added case, taken at $t = 31$ from a perspective (Figure 1a), top (Figure 1b) and side viewpoint (Figure 1c). In these figures field lines emerging below the surface (grey-scale map) are displayed in color. Figures 1d-1f show the time variations of emerged magnetic flux, free magnetic energy and relative magnetic helicity stored in the atmosphere $z \geq 0$, which are obtained by following the same procedure as described in Magara & Longcope (2003) and Magara (2012). The red vertical lines in these figures indicate $t = 31$.

We also derived these time series plots in all the other cases. We have then investigated mutual relationships among emerged magnetic flux, free magnetic energy and relative magnetic helicity during the dynamic evolution of an emerging magnetic field. Figure 2a shows a graph of emerged flux vs. free magnetic energy in a logarithmic scale while in Figures 2b and 2c log-log plots of emerged flux vs. relative magnetic helicity and relative magnetic helicity vs. free magnetic energy are presented. In these figures the red, blue, green, orange and violet lines represent that the twist parameter is given by 0.2, 0.35, 0.5, 0.8 and 1, respectively. A black dashed line in each figure indicates that the power law index is given by 2 (Figure 2a), 2 (Figure 2b) and 1 (Figure 2c). These indices are derived from the following rough estimates: $\Phi \sim BL^2 \propto B$, $\Delta E_m \sim B^2L^3 \propto B^2$ and $H_m \sim B^2L^4 \propto B^2$ where $\Phi$, $\Delta E_m$ and $H_m$ represent emerged flux, free magnetic energy and relative magnetic helicity while $B$ and $L$ are magnetic field strength and length, respectively.

3. Discussion

Figures 2a-2c show characteristics of emerged magnetic flux, relative magnetic helicity and free magnetic energy. In an ideal MHD system the first two quantities are conserved quantities, that is, as long as they are injected into the solar atmosphere through the emergence of a twisted flux tube, both of these quantities continuously increase there. On the other hand, during the dynamic evolution of an emerging magnetic field, free magnetic energy is continuously converted into the kinetic energy of expansion of an emerging field. This may be one of the reasons that a deviation from a power law

fitting line is observed in the plots of free magnetic energy vs. one of these conserved quantities.

In this respect, it might be possible to have a situation where the amount of free magnetic energy converted to kinetic energy is very small compared to the amount stored in an emerging flux region, especially in the core of an emerging flux region where a magnetic field shows quasi-static behavior (Magara & Longcope 2003). Let us consider a simple model of an emerging flux region based on a static field where no free magnetic energy is converted. We use a linear force-free field (LFFF) given by (e.g. Priest 1982)

$$B_x = -\sqrt{1 - \left(\frac{\alpha}{k}\right)^2} B_0 \cos(kx) e^{-\sqrt{k^2-\alpha^2}z}, \tag{1}$$

$$B_y = -\frac{\alpha}{k} B_0 \cos(kx) e^{-\sqrt{k^2-\alpha^2}z}, \tag{2}$$

and

$$B_z = B_0 \sin(kx) e^{-\sqrt{k^2-\alpha^2}z}, \tag{3}$$

where $k$ is the wave number while $\alpha$ and $B_0$ are constants representing the force-free parameter and field strength. An emerging flux region typically has the following evolutionary features (e.g. Shibata & Magara 2011 and references therein): i) at the beginning of emergence, the area of emerged flux at the surface is small and it gradually extends along a polarity inversion line (PIL) as emergence goes on. The direction of the PIL basically corresponds to the axis of an emerging flux tube. ii) at the beginning of emergence, the direction of an emerging field is almost perpendicular to a PIL, while it is gradually aligned with the PIL to form a sheared loop. To incorporate these basic features into the model, we assume a configuration illustrated in Figure 3a. By changing $k$ from $\infty$ toward $\alpha$, we make pseudo evolution of an emerging flux region, although the whole magnetic field is always in a static state. We then calculate emerged flux, relative magnetic helicity and free magnetic energy in the domain $\left(-\frac{2\pi}{k}, 0, 0\right) \leq (x, y, z) \leq \left(\frac{2\pi}{k}, \frac{\pi}{k}\frac{\alpha}{\sqrt{k^2-\alpha^2}}, \infty\right)$, given by (e.g. Berger 1985)

$$\Phi' = \frac{\alpha'}{k'^2\sqrt{k'^2 - \alpha'^2}}, \tag{4}$$

$$H'_m = \frac{\alpha'^2}{k'^4(k'^2 - \alpha'^2)}, \tag{5}$$

and

$$\Delta E'_m = \frac{\left(k' - \sqrt{k'^2 - \alpha'^2}\right)\alpha'}{k'^3(k'^2 - \alpha'^2)}, \tag{6}$$

respectively. Here $'$ means that a quantity is given in a dimensionless form. From Equations (4) and (5) we can immediately obtain $H'_m = \Phi'^2$, which is consistent with the simulations. A relation between $\Delta E'_m$ and $H'_m$ (or $\Phi'$) is a bit complicated; in the following we present an asymptotic result on this relation when $k' \to \infty$ and $k' \to \alpha'$:

$$\Delta E'_m \sim \frac{1}{2}\alpha' H'_m = \frac{1}{2}\alpha' \Phi'^2 \qquad \text{for } k' \to \infty \tag{7}$$

$$\Delta E'_m \sim \alpha' H'_m = \alpha' \Phi'^2 \qquad \text{for } k' \to \alpha' \qquad (8)$$

In Figure 3b the black line shows a log-log plot of $H'_m$ vs. $\Delta E'_m$ while red and blue dashed lines represent Equations (7) and (8), respectively. This suggests that power law indices for $(\Phi', \Delta E'_m)$-plot and $(H'_m, \Delta E'_m)$-plot are given by 2 and 1, which are the same values as obtained from the rough estimates in Section 2.

Comparing the simulations and the LFFF model raises a question about how much free magnetic energy remains in a magnetic structure evolving toward a well-developed state ready for producing an energy-release event. This question becomes further important when we use the scaling laws to estimate remaining free magnetic energy in an emerging flux region. To answer this question we need wide and detailed investigations of how the energy conversion of free magnetic energy proceeds in emerging flux regions of various magnetic configurations and how well force-free models give an estimate of free magnetic energy stored in these regions. For this purpose we will try to use a more general force-free field model than the LFFF model presented here and compare it to the simulations. Results of a detailed comparison will be reported in our forthcoming paper.

The author wishes to thank the Kyung Hee University for general support of this study. He also appreciates useful comments given by the referee. This study was financially supported by Basic Science Research Program (NRF-2013R1A1A2058705, PI: T. Magara) through the National Research Foundation of Korea (NRF) provided by the Ministry of Education, Science and Technology, as well as by the BK21 plus program through the NRF.

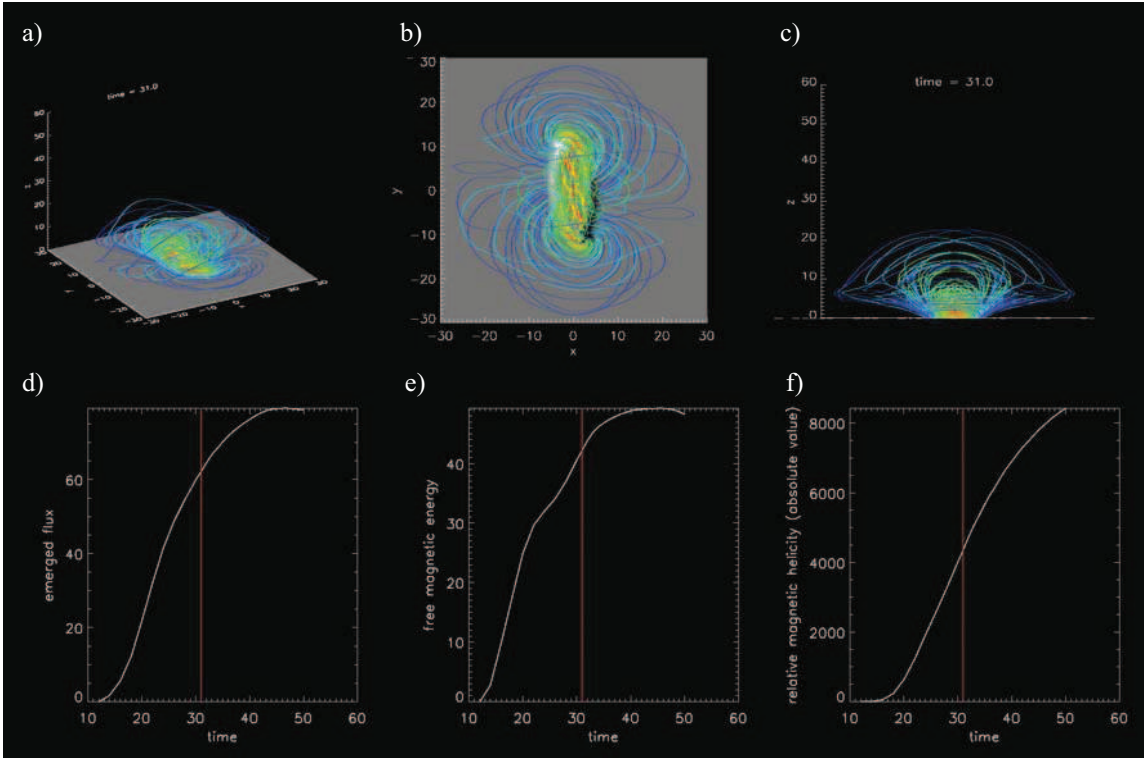

**Fig. 1.** Figures 1a-1c: Snapshots of an emerging magnetic field taken at $t = 31$ from a perspective (1a), top (1b) and side (1c) viewpoint are presented. The twist parameter is 0.8. Field lines are represented by colored lines and photospheric magnetic flux is given by a grey-scale map at $z = 0$. Figures 1d-1f: Time variations of emerged magnetic flux (1d), free magnetic energy (1e) and relative magnetic helicity (1f) are presented. The red vertical lines indicate $t = 31$.

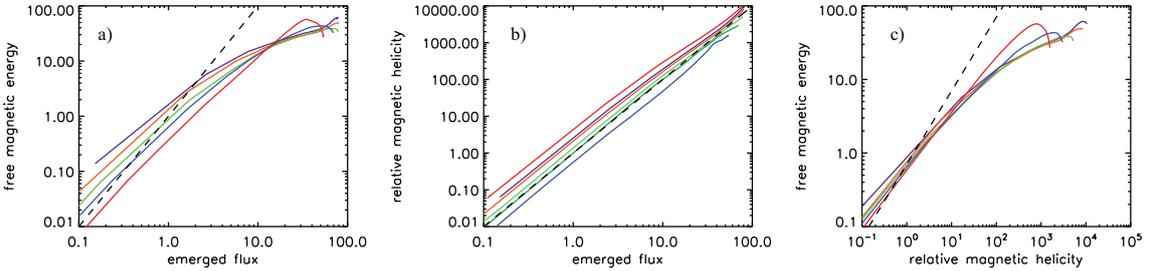

**Fig. 2.** (a) A log-log plot of emerged magnetic flux vs. free magnetic energy is presented. The twist parameter is 0.2 (red), 0.35 (blue), 0.5 (green), 0.8 (orange) and 1 (violet). The dashed line indicates power law index = 2. (b) Same as Figure 2a, except for a plot of emerged magnetic flux vs. relative magnetic helicity. The dashed line indicates power law index = 2. (c) Same as Figure 2a, except for a plot of relative magnetic helicity vs. free magnetic energy. The dashed line indicates power law index = 1.

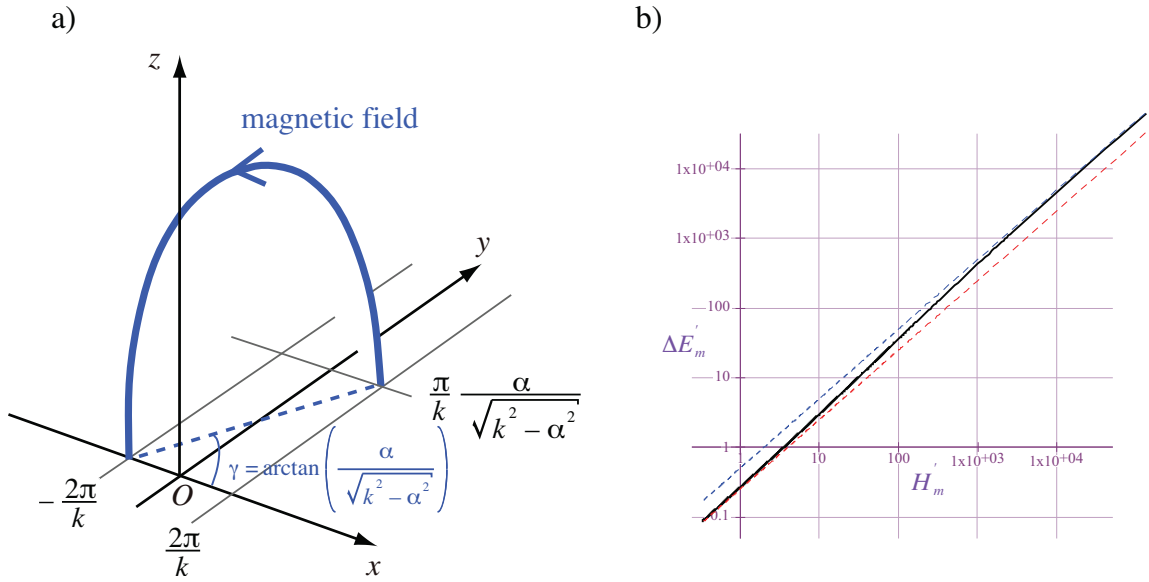

**Fig. 3.** (a) Schematic illustration of a 2.5-dimensional linear force-free field given by Equations (1) - (3) is presented. The $y$-axis represents a PIL while $\gamma$ is a shear angle. (b) A log-log plot of $H'_m$ vs. $\Delta E'_m$ given by Equations (5) and (6) is represented by a black line. The red and blue dashed lines show asymptotic lines given by Equations (7) and (8), respectively. Here $\alpha' = 0.5$.

8